\documentclass[%
 reprint,
 amsmath,amssymb,
 aps,
 prl,
]{revtex4-1}
\usepackage{cleveref}
\usepackage{graphicx}
\usepackage{subfigure}
\usepackage{dcolumn}
\usepackage{bm}

\begin{document}

\preprint{APS/123-QED}

\title{Bee-yond the Plateau: Training QNNs with Swarm Algorithms }

\author{Rub{\'e}n Dar{\'i}o Guerrero}
 \affiliation{Qauntum and Computational Chemistry Group, Universidad Nacional de Colombia - Bogot{\'a} Campus}
 \affiliation{NeuroTechNet S.A.S, 1108831, Bogotá, Colombia}

\date{\today}

\begin{abstract}
In the quest to harness the power of quantum computing, training quantum neural networks (QNNs) presents a formidable challenge. This study introduces an innovative approach, integrating the Bees Algorithm (BOA) to overcome one of the most significant hurdles—barren plateaus. Our experiments across varying qubit counts and circuit depths demonstrate the BOA's superior performance compared to the Adam algorithm. Notably, BOA achieves faster convergence, higher accuracy, and greater computational efficiency. This study confirms BOA's potential in enhancing the applicability of QNNs in complex quantum computations.
\end{abstract}

\maketitle





Quantum neural networks (QNNs) combine the power of quantum computing with the architecture of neural networks. By embedding quantum mechanics into neural networks, QNNs aim to significantly enhance information processing capabilities. These networks introduce quantum theoretical concepts and computing techniques into traditional neural networks, creating a new paradigm for computation \cite{10.5220/0004425701070112,schuld2015introduction,biamonte2017quantum}. In QNNs, each neuron's state is a coherent superposition of multiple quantum states, allowing for a more intricate non-local representation of information compared to classical neural networks \cite{10.5220/0004425701070112,schuld2015introduction,biamonte2017quantum}.

The potential of QNNs lies in their capacity to learn and predict data, akin to classical neural networks, but with the added advantage of utilizing quantum properties for increased computational power \cite{schuld2015introduction,biamonte2017quantum,10.3390/s23125683}. By merging classical neural network architectures with quantum computing principles, QNNs offer a unique approach to information processing that shows promise in addressing complex problems \cite{schuld2015introduction,biamonte2017quantum,boixo2018characterizing}. By leveraging quantum principles, these networks offer new avenues for addressing complex problems in quantum information science and beyond, potentially revolutionizing fields such as cryptography, materials science, and artificial intelligence.\cite{10.1088/1367-2630/11/11/113018}.

Expressivity is a key feature of QNNs, referring to their ability to represent and approximate a wide array of functions. Recent studies have drawn parallels between quantum neural networks and the universal approximation theorem fundamental to the expressivity of classical neural networks \cite{10.1103/physrevresearch.3.l032049}. This comparison suggests that QNNs can effectively approximate the behavior of quantum systems with high accuracy, showcasing their computational power and versatility.

The architecture of QNNs is meticulously designed with quantum neurons arranged in specific topological structures, highlighting the significance of quantum entanglement and network connectivity in shaping the computational capabilities of these systems \cite{10.1002/que2.75}. The incorporation of quantum gates as neural network weights further enhances the adaptability and learning potential of QNNs, enabling them to tackle tasks that may pose challenges for classical neural networks. 

Researchers have explored the application of QNNs in diverse domains such as control theory and servo systems, underscoring the versatility of these networks in various computational tasks \cite{10.3233/his-130174}. By employing qubit neurons as information processing units, QNNs have been instrumental in designing adaptive controllers and learning algorithms, showcasing the potential of quantum-enhanced computing in practical applications \cite{10.3233/his-130174}.

In addition to their computational prowess, QNNs offer advantages such as exponential memory capacity, accelerated learning, and the prevention of catastrophic forgetting, a phenomenon observed in classical neural networks due to pattern interference \cite{jiang2022quantum}. These attributes position QNNs as a compelling area of research for advancing machine learning algorithms and with the potential of finding new ways for driving forward quantum information technology.

Ongoing efforts have shown that a class of QNNs achieves a considerably better effective dimension than comparable feedforward networks and trains faster, suggesting an advantage for quantum machine learning \cite{10.48550/arxiv.2011.00027}. However, a significant challenge in training QNNs is the presence of barren plateaus, where the gradients of the cost function become exponentially vanishing, hindering effective optimization \cite{10.1038/s41467-018-07090-4}. Barren plateaus have been observed in various quantum neural network architectures, representing a fundamental obstacle to the successful training of QNNs and limiting their practical applicability \cite{10.48550/arxiv.2303.03227}.

Researchers have explored various strategies to address the barren plateau problem in QNN training. One approach involves task-specific initialization to circumvent barren plateaus and enhance the efficiency of training quantum circuits \cite{10.21203/rs.3.rs-2139932/v1}. Additionally, layerwise training has been proposed, where a shallow quantum circuit is initially trained before gradually adding layers to alleviate the barren plateau issue \cite{10.48550/arxiv.2108.02792}. Moreover, introducing entanglement constraints has been suggested as a way to mitigate barren plateaus and improve the learning process in QNNs \cite{10.48550/arxiv.2012.12658}.

In the realm of quantum computing, swarm algorithms have shown promise in optimizing quantum-inspired models. \citet{10.3390/a12070138} proposed a quantum-behaved neurodynamic swarm optimization approach combined with a neural network and the QPSO algorithm to tackle nonconvex optimization problems with constraints, showcasing the potential of swarm intelligence in quantum computing applications \cite{10.3390/a12070138}. Additionally, \citet{10.14257/ijsh.2016.10.6.18} investigated a quantum adaptive PSO algorithm for optimizing radial basis function neural networks, highlighting the synergy between quantum-inspired algorithms and neural network optimization \cite{10.14257/ijsh.2016.10.6.18}. These studies underscore the growing interest in leveraging swarm algorithms for enhancing quantum computing tasks, including quantum neural network training.

The Bees optimization algorithm (BOA) is a nature-inspired optimization technique that mimics the foraging behavior of honey bees to solve complex optimization problems \citep{10.1243/09544062jmes1494}. This algorithm has shown significant strengths in its search mechanism, enabling it to outperform other swarm algorithms, especially in addressing hard cost functions characterized by barren plateaus.

One of the key strengths of BOA lies in its ability to accurately search narrow fitness valleys and holes, as well as optimize highly multimodal functions \citep{10.1177/0959651811422759}. This precision in searching allows BOA to navigate challenging landscapes efficiently, making it particularly effective in finding optimal solutions in complex optimization scenarios. Additionally, BOA has been observed to excel in position and balance control applications, outperforming traditional methods and showcasing superior results \citep{10.21541/apjess.1084875}.

Empirical studies have shed light on the properties of BOA and related swarm optimization techniques (SOT), emphasizing how parameterization significantly impacts the search effort and overall performance \citep{10.1016/j.swevo.2020.100746}. By fine-tuning parameters, BOA can adapt its search strategy to suit different optimization landscapes, enhancing its effectiveness in exploring solution spaces efficiently.

In recent years, the optimization of QNNs has faced significant challenges, notably the problem of barren plateaus. Barren plateaus are regions in the parameter space where the gradient of the loss function becomes exponentially small, making training inefficient and often unsuccessful. Addressing this issue is crucial for the practical deployment of QNNs in complex computational tasks.

In this letter, we present a novel approach that integrates BOA into the training process of QNNs to effectively tackle the barren plateau problem.  By incorporating BOA, we demonstrate substantial improvements in the optimization of QNN parameters, leading to enhanced training efficiency and model performance. Our results indicate that this method not only mitigates the challenges posed by barren plateaus but also provides a more reliable path to finding the global minimum in the optimization landscape. This advancement significantly enhances the learning capabilities of QNNs, making them more feasible for a wide range of applications in quantum informations sciences.

To illustrate the practical applicability of our approach, we focus on the variational quantum eigensolver (VQE) algorithm \cite{peruzzo2014variational,bravo2020scaling,consiglio2022variational} . The VQE algorithm is a hybrid quantum-classical algorithm used to find the ground state energy of a quantum system. Specifically, we apply BOA to minimize the loss function

\begin{equation}\label{eq:VQE}
E(\boldsymbol{\theta})=\left\langle\psi(\boldsymbol{\theta})\left|H_{\mathrm{TFIM}}\right| \psi(\boldsymbol{\theta})\right\rangle,
\end{equation}

and determine the ground state of the transverse field Ising model (TFIM) Hamiltonian $H_{\text{TFIM}}$. Here, $|\psi(\boldsymbol{\theta})\rangle = U(\boldsymbol{\theta})|+\rangle^{\otimes n}$, and the Hamiltonian is given by

\begin{equation}
H_{\text{TFIM}} = -\sum_{i=1}^{n_f} \sigma_i^z \sigma_{i+1}^z - h \sum_{i=1}^n \sigma_i^x,
\end{equation}

where $\sigma_i^\mu$ denotes the Pauli matrices acting on qubit $i$, and $h$ is the strength of the transverse field. We set $h=1.1$ and consider both open ($n_f = n-1$) and closed ($n_f = n$) boundary conditions, with $\sigma_{n+1}^\mu = \sigma_1^\mu$ in the latter.

We employ a Hamiltonian variational ansatz for the QNN, where the QNN is represented as a parameterized quantum circuit $U(\boldsymbol{\theta})$ that acts on quantum states in the training set as $U(\boldsymbol{\theta})|\psi_\mu\rangle$ \cite{PhysRevA.92.042303,PRXQuantum.1.020319} . The ansatz has an $L$-layered periodic structure given by

\begin{equation}\label{eq:anzats}
U(\boldsymbol{\theta}) = \prod_{l=1}^L U_l(\boldsymbol{\theta}_l), \quad U_l(\boldsymbol{\theta}_l) = \prod_{k=1}^K e^{-i \theta_{lk} H_k},
\end{equation}

where $l$ indicates the layer, $k$ spans the traceless Hermitian operators $H_k$, and $\boldsymbol{\theta}_l = (\theta_{l1}, \ldots, \theta_{lK})$ are the parameters within a single layer. The total set of trainable parameters is $\boldsymbol{\theta} = \{\boldsymbol{\theta}_1, \ldots, \boldsymbol{\theta}_L\}$.

 As highlighted in \cite{larocca2022diagnosing}, \cref{eq:anzats} encompasses several specific ansatzes, including the hardware-efficient ansatz \cite{kandala2017hardware}, the quantum alternating operator ansatz (QAOA) \cite{farhi2014quantum,hadfield2019quantum}, Adaptive QAOA \cite{PhysRevResearch.4.033029}, the Hamiltonian Variational Ansatz (HVA) \cite{wecker2015progress}, and the Quantum Optimal Control Ansatz \cite{PhysRevResearch.3.023092}, among others \cite{PhysRevA.104.032401}.

We investigate the set of reachable states through the action of $U(\boldsymbol{\theta})$ on $\left|\psi_\mu\right\rangle$. Given a set of parameters $\boldsymbol{\theta}$ and an infinitesimal perturbation $\boldsymbol{\delta}$, presumably obtained from an update rule, we quantify the distance $\mathcal{D}$ between the quantum states $\left|\psi_\mu(\boldsymbol{\theta})\right\rangle=U(\boldsymbol{\theta})\left|\psi_\mu\right\rangle$ and $\left|\psi_\mu(\boldsymbol{\theta}+\boldsymbol{\delta})\right\rangle=U(\boldsymbol{\theta}+\boldsymbol{\delta})\left|\psi_\mu\right\rangle$. The second-order Taylor expansion of $\mathcal{D}$ is given by the Fubini-Study metric \cite{cheng2010quantum,meyer2021fisher} as:

\begin{equation}
\mathcal{D}\left(\left|\psi_\mu(\boldsymbol{\theta})\right\rangle,\left|\psi_\mu(\boldsymbol{\theta}+\boldsymbol{\delta})\right\rangle\right)=\frac{1}{2} \boldsymbol{\delta}^T \cdot F_\mu(\boldsymbol{\theta}) \cdot \boldsymbol{\delta}.
\end{equation}

Here, $F_\mu(\boldsymbol{\theta})$ is the Quantum Fisher Information Matrix (QFIM) for the state $\left|\psi_\mu\right\rangle$. As described in \cite{liu2020quantum}, the QFIM is an $M \times M$ matrix with elements:
\begin{equation}
\begin{aligned}
{\left[F_\mu(\boldsymbol{\theta})\right]_{i j}=4 \operatorname{Re}[} & \left\langle\partial_i \psi_\mu(\boldsymbol{\theta}) \mid \partial_j \psi_\mu(\boldsymbol{\theta})\right\rangle \\
& \left.-\left\langle\partial_i \psi_\mu(\boldsymbol{\theta}) \mid \psi_\mu(\boldsymbol{\theta})\right\rangle\left\langle\psi_\mu(\boldsymbol{\theta}) \mid \partial_j \psi_\mu(\boldsymbol{\theta})\right\rangle\right],
\end{aligned}
\end{equation}
where $\left|\partial_i \psi_\mu(\boldsymbol{\theta})\right\rangle = \partial\left|\psi_\mu(\boldsymbol{\theta})\right\rangle / \partial \theta_i = \partial_i\left|\psi_\mu(\boldsymbol{\theta})\right\rangle$ for $\theta_i \in \boldsymbol{\theta}$. The QFIM plays a crucial role in imaginary time evolution algorithms \cite{mcardle2019variational} and in quantum-aware optimizers like quantum natural gradient descent \cite{stokes2020quantum,gacon2021simultaneous,koczor2022quantum,haug2022natural}. The rank of the QFIM indicates the number of independent directions in state space that can be explored with infinitesimal changes $\boldsymbol{\delta}$ in $\boldsymbol{\theta}$.

We then consider the formulation of the loss function, which is crucial for understanding the optimization landscape:
\begin{equation} \label{eq:Loss}
\mathcal{L}(\boldsymbol{\theta})=\sum_{\left|\psi_\mu\right\rangle \in \mathcal{S}} c_\mu \operatorname{Tr}\left[U(\boldsymbol{\theta})\left|\psi_\mu\right\rangle\left\langle\psi_\mu\right| U^{\dagger}(\boldsymbol{\theta}) O\right],    
\end{equation}
where $c_\mu$ are real coefficients associated with each state $\left|\psi_\mu\right\rangle$ in the set $\mathcal{S}$, the training space, and $O$ is a Hermitian operator.

Further exploring the optimization landscape, we analyze the curvature of the loss function through the Hessian matrix $\nabla^2 \mathcal{L}(\boldsymbol{\theta})$, an $M \times M$ matrix defined as:
\begin{equation}
\left[\nabla^2 \mathcal{L}(\boldsymbol{\theta})\right]_{ij} = \partial_i \partial_j \mathcal{L}(\boldsymbol{\theta}).
\end{equation}
Evaluating the gradient and Hessian at a given point provides a quadratic approximation of the loss function. The Hessian's eigenvectors corresponding to positive (negative) eigenvalues indicate directions of positive (negative) curvature. The rank of $\nabla^2 \mathcal{L}(\boldsymbol{\theta})$ signifies the number of directions causing second-order changes in the loss, with zero eigenvalues pointing to barren plateaus. The Hessian has been employed to characterize the loss landscapes of variational quantum algorithms \cite{huembeli2021characterizing,kim2022quantum,dalgaard2022predicting,dalgaard2020hessian}.

\begin{figure}[htbp]
    \centering
    \subfigure[Training curves for the VQE implementation using the Adam optimization algorithm, illustrating a typical optimization stall caused by barren plateaus.]{
        \includegraphics[width=0.5\textwidth]{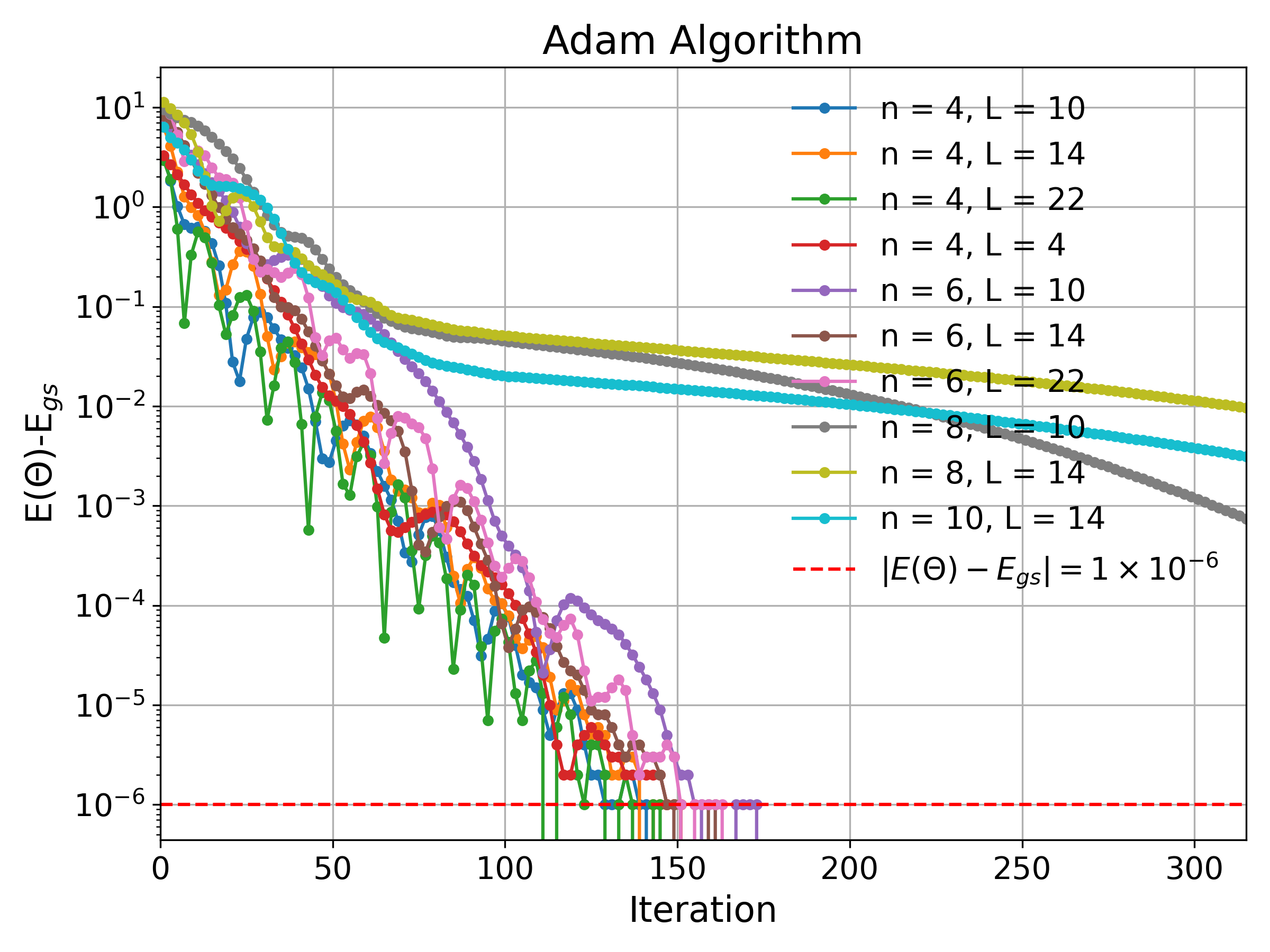}
        \label{fig:Adam}
    }
    \subfigure[Training curves for VQE implementation using BOA]{
        \includegraphics[width=0.5\textwidth]{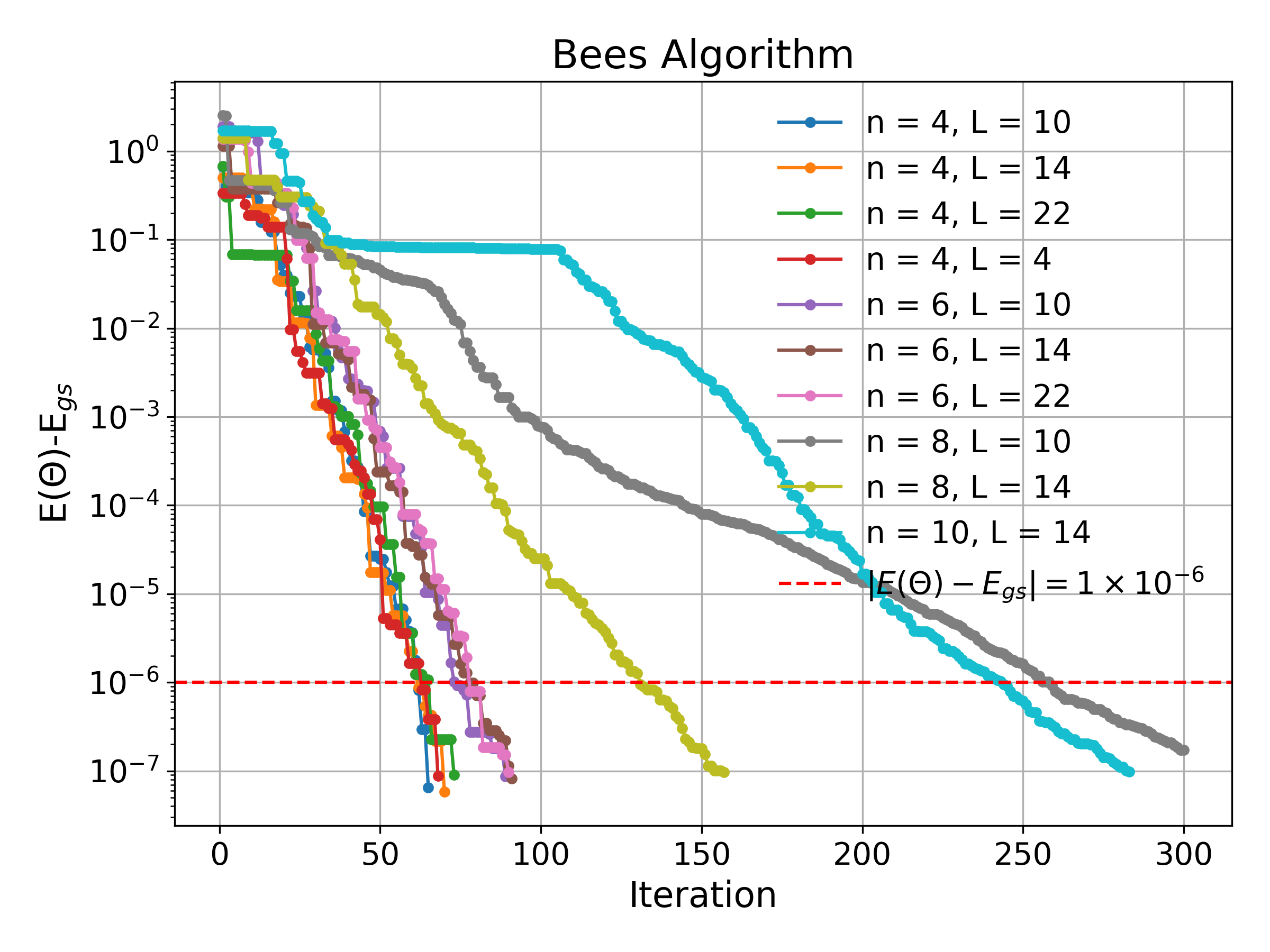}
        \label{fig:BOA}
    }
    \caption{
    Convergence of $|E(\Theta) - E_{gs}|$ versus iteration number. A Hamiltonian variational ansatz with periodic boundary conditions was employed to address the VQE task as described in \cref{eq:VQE}. The dashed horizontal line indicates the accuracy goal. Panel (a) displays the best performance of the Adam algorithm across 30 realizations, each starting from different random initial conditions. Panel (b) presents the results obtained using the BOA with a single swarm.
    }
    \label{fig:VQE_results}
\end{figure}

To establish a reliable benchmark for comparison with BOA, we will employ the Adaptive Moment Estimation (Adam) algorithm, which is extensively utilized in neural network training due to its multiple practical benefits. Adam integrates the advantages of two other stochastic gradient descent extensions: the Adaptive Gradient Algorithm (AdaGrad) and Root Mean Square Propagation (RMSProp). This integration enables Adam to effectively handle sparse gradients on noisy problems, making it particularly advantageous for large datasets and high-dimensional parameter spaces. Additionally, Adam automates the adjustment of learning rates by computing adaptive rates for each parameter. This capability not only promotes faster convergence but also reduces the necessity for manual tuning of the learning rate, thereby enhancing user-friendliness and efficiency across a broad spectrum of applications. Moreover, Adam's robustness to initial configurations and hyperparameter settings further contributes to its widespread adoption among machine learning practitioners and have being used by previous works in the training of VQE.\cite{larocca2023theory}

Figure \ref{fig:VQE_results} demonstrates the performance of minimizing the loss function as detailed in \cref{eq:VQE} across QNNs with varying qubit counts of $n=$ 4, 6, 8, and 10 and circuit depths $L=$ 4, 10, 14, and 22. These depths are sufficiently large to ensure overparametrization, which is necessary for making the QNNs trainable. Panel (a) of the figure illustrates the best performance achieved by the Adam algorithm across 30 realizations, each initialized with different random conditions. Panel (b) details the results obtained using the Bees Algorithm (BOA) employing a single swarm. This setup includes $n_s=10$ scout bees, with $n_b=5$ sites selected from the $n_s$ visited sites, $n_e=1$ elite site among the $n_b$ selected, $n_{re}=15$ bees recruited to forage at the elite site, and $n_{rb}=10$ bees recruited for the other $n_b-n_e$ selected sites. Additionally, the heuristics include a setting where a site is abandoned after $st_{lim}=10$ cycles.

The subsequent performance comparison between the Adam algorithm and the Bees Algorithm, as illustrated in \cref{fig:VQE_results}, reveals distinct differences across several key metrics. Notably, the Bees Algorithm demonstrates significantly faster convergence, achieving the target accuracy of \( |E(\Theta) - E_{gs}| \leq 10^{-6} \) within about 50 iterations for configurations up to $n=6$ and within 300 iterations for $n>6$. In contrast, the Adam algorithm often requires two to three times more iterations to reach comparable accuracy, and in some cases with $n=8$ and $n=10$ at certain depths, it fails to achieve the desired accuracy within 300 iterations. The Bees Algorithm consistently outperforms in computational efficiency, requiring fewer iterations to achieve optimal results, making it ideal for resource-constrained scenarios.



In evaluating the robustness and stability, the Bees Algorithm shows remarkable consistency and reliability in converging to the accuracy goal without significant fluctuations, whereas the Adam algorithm exhibits less predictable behavior, with oscillations and failures, particularly in more complex configurations. These findings suggest that the Bees Algorithm offers superior performance in terms of convergence speed, accuracy, computational efficiency, robustness, and stability, making it the recommended choice for this computational problem. While the Adam algorithm might still be suitable for less complex scenarios, for high-demand applications requiring precise and efficient outcomes, the Bees Algorithm clearly outperforms.

See Supplemental Material at [URL will be inserted by publisher] for detailed characterization of the loss landscape using QFIM and its Hessian, performed with BOA. This analysis provides deeper insights into the optimization dynamics and complexities in QNNs, complementing our findings and offering a more comprehensive understanding of the factors influencing QNN training.

By presenting this novel integration of BOA into the training of QNNs, we open new avenues for addressing fundamental challenges in quantum machine learning, with potential applications extending to various areas of physics and beyond.

\bibliography{apssamp}

\end{document}